\documentclass[11pt,twoside]{article}  
\usepackage{adassconf}

\begin{document}   

%
%
%

\paperID{O1-3}

%
%
%
%

\title{Palomar-QUEST: A case study in designing sky surveys in the VO era}
\titlemark{The Palomar-QUEST survey}

%
%
%

\author{Matthew J. Graham, Roy Williams, S. G. Djorgovski, Ashish Mahabal}
\affil{California Institute of Technology, Pasadena, CA 91125, USA}
\author{Charles Baltay, Dave Rabinowitz, Anne Bauer, Jeff Snyder, Nick Morgan, Peter Andrews}
\affil{Yale University, New Haven, CT 06511, USA}
\author{Alexander S.  Szalay}
\affil{Dept. Physics and Astronomy, The Johns Hopkins University, Baltimore, MD 21218, USA}
\author{Robert J. Brunner}
\affil{Dept. of Astronomy \& NCSA, University of Illinois, Urbana, IL 61801, USA}
\author{Jim Musser}
\affil{Indiana University, Bloomington, Indiana, USA}

%
%

\contact{Matthew Graham }
\email{mjg@astro.caltech.edu }

%
%
%
%
%

\paindex{Graham, M. J.}
\aindex{Williams, R.}     
\aindex{Djorgovski, S. G.}
\aindex{Mahabal, A.}
\aindex{Baltay, C.}
\aindex{Rabinowitz, D.}
\aindex{Bauer, A.}
\aindex{Snyder, J.}
\aindex{Morgan, N.}
\aindex{Andrews, P.}
\aindex{Szalay, A. S.}
\aindex{Brunner, R. J.}
\aindex{Musser, J.}

%
%

\authormark{Graham et al.}

%
%

\keywords{ synoptic surveys, multicolour surveys, wide area surveys, VO }


\begin{abstract}          
The advent of wide-area multicolour synoptic sky surveys is leading to data sets unprecedented in size, complexity and data throughput. VO technology offers a way to exploit these to the full but requires changes in design philosophy.
The Palomar-QUEST survey is a major new survey being undertaken by Caltech, Yale, JPL and Indiana University to repeatedly observe $\frac{1}{3}$ of the sky ($\sim 15000$ sq. deg. between $-27^\circ \le \delta \le 27^\circ$) in seven passbands. Utilising the 48-inch Oschin Schmidt Telescope at the Palomar Observatory with the 112-CCD QUEST camera covering the full 4$^\circ$ x 4$^\circ$ field of view, it will generate $\sim 1$TB of data per month.
In this paper, we review the design of QUEST as a VO resource, a federated data set and an exemplar of VO standards.
\end{abstract}

%
%

\section{A new era in astronomy}

The new availability of wide-field images from Schmidt telescopes in the 1940's meant that astronomers no longer had to make educated guesses about where to look to find new and interesting phenomena but were now spoilt for choice. The advent of synoptic surveys presents more extreme opportunities; as an illustration, consider the SDSS which over the course of 5 years represents a factor of a million increase in information over previous surveys; however, the \htmladdnormallinkfoot{LSST}{http://www.lssto.org} (Large Sky Synoptic Telescope, Tyson (2002)) will amass a SDSS every 3 nights. 

Although overviews of synoptic surveys are riddled with cliches concerning undiscovered countries and uncharted waters, the exploration of the temporal domain results in data sets that are not just more voluminous than before, but far richer and more complex (Paczynski 2001; Djorgovski et al. 2000). This presents challenges to all aspects of astronomy: data gathering, distribution, reduction, analysis, storage, archiving, dissemination and mining. VO technologies are being designed precisely to meet these types of challenges, but to use them requires changes in survey design philosophies. 

\section{The Palomar-QUEST survey}

The \htmladdnormallinkfoot{Palomar-QUEST survey}{http://hepwww.physics.yale.edu/quest/palomar.html; http://www.astro.caltech.edu/~aam/science/quest}  is a major new survey being undertaken by Caltech, Yale, JPL and Indiana University employing the world's largest astronomical camera and the recently refurbished  Oschin Schmidt telescope at Palomar to observe a third of the sky ($\sim 15000$ sq. deg. between $-27^\circ \le \delta \le 27^\circ$) a minimum of 8 times in 7 passbands to nominally twice the depth of SDSS. 

The QUEST camera consists of 112 CCDs arranged in four filter strips. Each CCD has 2400 $\times$ 600  13$\mu$m $\times$ 13$\mu$m pixels, giving a total of 161 $\times$ 10$^6$ pixels. At the prime focus of the Oschin Schmidt, QUEST covers a sky area of 4.6$^\circ \times$ 3.6$^\circ$ (the effective area is $\sim 10$ sq. deg) and in a night can survey $\sim 500$ sq. deg. Two filter sets are used: Johnson $UBRI$ and Gunn $riz$, with a doubling of Gunn $z$ to afford extra depth. 

The data rate is 2.45MB/s and with a monthly average of 10 nights' observing, QUEST produces $\sim 1$TB of data/month.

Some of the immediate science goals are searching for high redshift quasars, strong gravitational lensing, supernovae and GRBs, and near-Earth asteroids and trans-Neptunian objects. Obviously once there is a sufficient body of repeat observations, searching for new types of variable object and phenomena will play a dominant part; in particular, a rapid response mechanism to transients (see section 4) is planned.

\section{QUEST and VO technologies}

As this survey is one of the first of the new breed of synoptic surveys, it is being used as a testbed for the VO technologies which will enable astronomers to exploit such surveys to the full.
There are currently four areas of attention:

\subsection{Data distribution}

Different groups want to process the raw data in different ways to optimize the detection of specific types of object. Access requirements to the data are also either near real-time or delayed. Data distribution must be secure, fault tolerant (error checking, multiply redundant) and accountable (transaction logging).

\subsection{Data processing}

The nature of the data is extremely well suited to parallelization, either on a multi-processor machine or in a more general distributed computing environment, e.g. an advanced highly CPU-intensive pipeline would be a suitable Grid-level application.

\subsection{Data analysis}

The identification of variable objects poses many problems:
\begin{itemize}
\item associating different observations under different conditions (e.g. seeing) with the same identification;
\item handling objects which only appear once (e.g. supernovae)
\item handling moving objects (e.g. asteroids)
\item optimally characterizing the variability of an object (periodic/aperiodic)
\item determining the best sampling strategy to maximize the range of temporal baselines covered
\end{itemize}

Other federated data sets will be employed in the data analysis to assist identification, e.g. SDSS, DPOSS, 2MASS.

\subsection{Data dissemination}

The deployment of QUEST as a federated data set needs to support both interactive and batch mode access.  Access to data products also needs to be transparent to the access rights of different users: QUEST survey team, collaborators and the general astronomy community.

\section{Integration example: 4 minute alert of detection of transient events}

To illustrate how QUEST will make use of VO technologies in an integrated fashion, consider one of the pipeline systems under construction (see Fig. 1 for a cartoon depiction): this will produce real time (within four minutes of the data being taken) alerts of transient events (e.g. supernovae). The specific processes which need to mesh are:
\begin{itemize}
\item {\em Distribution:} Every 140s, 112 $\times$ 3.1MB raw fits files are produced at Palomar and streamed to Caltech (at 10MB/s) where the CIT Data Broker distributes the data to other sites, the raw image archive and the reduction pipeline
\item {\em Processing:} The CIT Fast Pipeline computes a real-time flat and extracts objects - each field produces $\sim$10000 objects
\item {\em Analysis:} Variable and transient objects are detected by comparing the latest observations with the fiducial sky (composed from all QUEST observations and possibly other data sets) in the master archive. They are processed to determine whether they might be asteroids and checked against lists of known variables. Source classification is attempted using other federated data archives.
\item {\em Dissemination:} The Alert Decision Engine decides whether an alert should be issued based on decision algorithms and all available data and posts results to the website.
\end{itemize}

\begin{figure}
\plotone{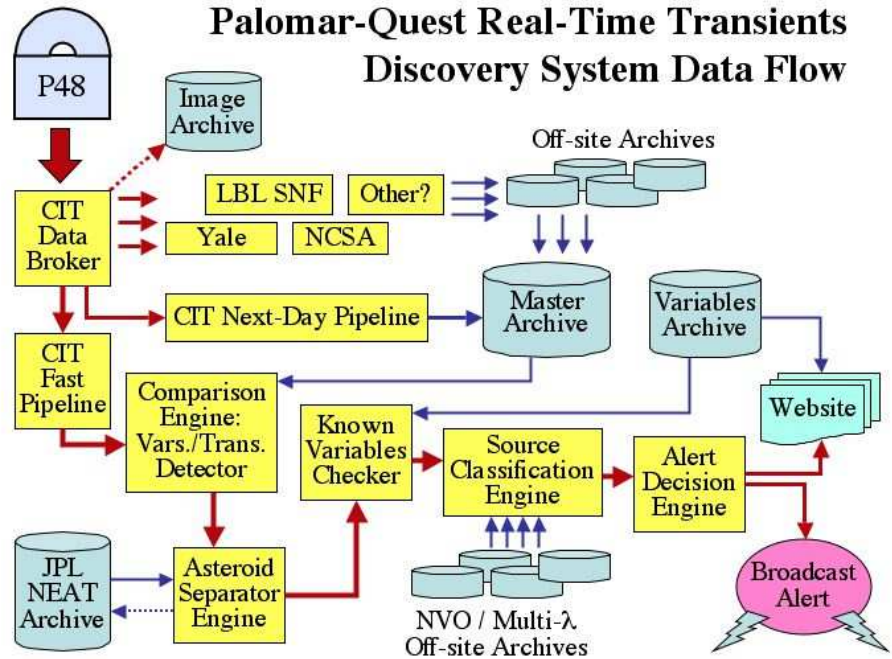}
\caption{Cartoon of the real time variable object detection pipeline which will issue an alert within four minutes of detection.}
\end{figure} 

\section {Conclusions}
Palomar-QUEST is the prototype VO-integrated synoptic sky survey and marks the beginning of an exciting new era in astronomy: the characterization of the variable optical sky.

\acknowledgements
MJG, RW, SGD and AAM acknowledge support from the NSF NVO program.
SGD, AAM and MJG also acknowledge a partial support from the NASA AISRP program.


\begin{references}
\reference Djorgovski, S.\ G.\  et al. 2001, in  Virtual Observatories of the Future, ed.\ R.\ J.\ Brunner, S.\ G.\  Djorgovski \& A.\ S.\ Szalay (ASP Conf. Ser. 225), 52  
\reference Paczynski, B.\ 2000, \pasp, 112, 1281
\reference Tyson, J.\ A.\  2002, SPIE, 4836, 10 
\end{references}
\end{document}